\documentstyle[12pt,amsbsy,epsf]{article}
\begin{document}

\setlength{\textwidth}{166mm} \setlength{\oddsidemargin}{0mm}
\setlength{\evensidemargin}{0mm} \setlength{\headheight}{0mm}
\setlength{\topmargin}{-5mm} \setlength{\textheight}{222mm}

\newcommand{\beq}{\begin{equation}}
\newcommand{\eeq}{\end{equation}}
\newcommand{\eq}[1]{(\ref{#1})}
\newcommand{\beqn}{\begin{eqnarray}}
\newcommand{\eeqn}{\end{eqnarray}}
\newcommand{\dst}{&\displaystyle}
\newcommand{\ba}{\bar \alpha}
\newcommand{\bb}{\bar \beta}
\newcommand{\fr}[2]{\frac{#1}{#2}}
\newcommand{\p}{\mbox{${\bf p}$}}
\newcommand{\q}{\mbox{${\bf q}$}}
\newcommand{\vj}{\mbox{${\bf j}$}}
\newcommand{\r}{\mbox{${\bf r}$}}
\newcommand{\R}{\mbox{${\bf R}$}}
\newcommand{\n}{\mbox{${\bf n}$}}
\newcommand{\bk}{\mbox{${\bf k}$}}
\newcommand{\bv}{\mbox{${\bf v}$}}
\newcommand{\s}{\mbox{${\bf s}$}}
\newcommand{\bS}{\mbox{${\bf S}$}}
\newcommand{\bd}{\mbox{${\bf d}$}}
\newcommand{\bu}{\mbox{${\bf u}$}}
\newcommand{\E}{\mbox{${\bf E}$}}
\newcommand{\A}{\mbox{${\bf A}$}}
\newcommand{\B}{\mbox{${\bf B}$}}
\newcommand{\e}{\mbox{${\bf e}$}}
\newcommand{\f}{\mbox{${\bf f}$}}
\newcommand{\va}{\mbox{${\bf a}$}}
\newcommand{\vq}{\mbox{${\bf q}$}}
\newcommand{\vI}{\mbox{${\bf I}$}}
\newcommand{\ep}{\mbox{${\varepsilon}$}}
\newcommand{\al}{\mbox{${\alpha}$}}

\begin{titlepage}
\vspace{1cm}

\begin{center}
{\large \bf  P and T odd electromagnetic moments of deuteron\\
in chiral limit}
\end{center}

\begin{center}
I.B. Khriplovich\footnote{khriplovich@inp.nsk.su} and R.V. Korkin
\end{center}
\begin{center}
Budker Institute of Nuclear Physics\\
630090 Novosibirsk, Russia,\\
and Novosibirsk University
\end{center}

\bigskip

\begin{abstract}
P odd anapole moment of the deuteron is found in the chiral limit, 
$m_{\pi} \to 0$. 
The contact current generated by the P odd pion exchange does not contribute 
to the deuteron anapole.
Being combined with usual radiative corrections
to the weak electron -- deuteron interaction, our calculation results in a
sufficiently accurate theoretical prediction for the corresponding 
effective constant $C_{2d}$. The experimental measurement of this
constant would give 
valuable information on the P odd $\pi$NN constant and on the $s$-quark
content of nucleons. We calculate also in the same limit $m_{\pi} \to 0$ 
the deuteron P odd and T odd multipoles: electric 
dipole moment and magnetic quadrupole moment.
\end{abstract}


\vspace{7cm}

\end{titlepage}

\begin{flushright}
{\it Dedicated to the memory of Slava Ryndin}
\end{flushright}

\bigskip
\bigskip

\section{Introduction} The investigations of nuclear parity violation,
both theoretical and experimental, have already a long history. New 
light on this problem is shed by observation of the 
nuclear anapole moment (AM) of $^{133}$Cs in atomic experiment~\cite{wi}. 
The result of this  
experiment is in a reasonable quantitative agreement with the
theoretical predictions, starting with~\cite{fk,fks}, if the so-called 
``best values''~\cite{ddh} are chosen for the parameters of P odd 
nuclear forces. 

The AM is a rather peculiar multipole in the following sense
(for a more detailed discussion see, for instance,~\cite{kh}). The
interaction of a charged probe particle with an anapole moment is of
a contact nature. Therefore, for instance, the interaction of the 
electron with the nucleon AM, being on the order of $\al G$, cannot
be distinguished in general case from other electromagnetic radiative 
corrections to the weak electron-nucleon interaction due to the
neutral currents. And in a gauge theory of electroweak interactions
only the total scattering amplitude, i.e., the sum of all diagrams on 
the order of $\al G$, is gauge-invariant, independent of the gauge 
choice for the Green's functions of heavy vector bosons (here 
$\al$ is the fine-structure constant, $G$ is the Fermi weak 
interaction constant). No wonder that, generally speaking, the AM of
an elementary particle or a nucleus is not gauge-invariant, i.e.,
physically well-defined, quantity. However, there is a special case
where the anapole moment has a real independent physical meaning. 
In heavy nuclei, of course $^{133}$Cs 
included, the AM is enhanced $\sim A^{2/3}$~\cite{fks} ($A$ is 
the atomic number), as distinct from common radiative corrections. 
By the way, it means that there is an intrinsic limit for the
relative accuracy, $\sim A^{-2/3}$, with which the AM of a heavy 
nucleus can be defined at all. For $^{133}$Cs this limiting
accuracy is about 4\%.

There is one more object, the deuteron, whose anapole moment could make
sense for a sufficiently large P odd $\pi$NN constant~\cite{fk}. 
The deuteron is a loosely bound system of a relatively 
simple structure. Therefore, there are all the 
reasons to believe that its AM is induced mainly by the P odd
$\pi$-meson exchange, pion being the lightest possible mediator of the
nucleon-nucleon weak interaction. The problem of the deuteron AM was 
discussed phenomenologically in~[2,6-8]. 
In the present work the deuteron AM, as induced by the P odd
$\pi$-meson exchange, is explicitly expressed through the P odd 
$\pi$NN coupling constant. The same problem was considered in recent 
paper~\cite{ss}. The result of the original version of~\cite{ss} was
much smaller than ours because a leading contribution, that of the 
isovector magnetic moment of the nucleon, was omitted in it. After 
acquaintance with the preprint of the 
present work, the authors of~\cite{ss} corrected their result (see their
revised preprint~\cite{ss}, Section VI. Erratum and Addendum). On the 
other hand, under the influence of~\cite{ss}, we have refined our own 
calculations with the results presented below. 

The obtained result for the deuteron AM is singular as $1/m_{\pi}$ in the
limit $m_{\pi} \to 0$ (of course, when going over to this limit, one 
should keep the deuteron radius larger than the Compton wave length of the 
pion). Being combined with the radiative corrections to the weak 
electron -- deuteron scattering amplitude~\cite{ms}, which are regular in
$m_{\pi}$, our calculations result
in a sufficiently accurate value for the corresponding effective constant 
$C_{2d}$.
 
We also calculate here the P odd and T odd electromagnetic moments of the 
deuteron.

\bigskip

\section{The deuteron anapole moment}

It is convenient to start the discussion with the nucleon AM in the
chiral limit. It was shown in 1980 by A.I. Vainshtein and one of the 
authors (I.Kh.) to be given in this limit by the diagrams 1 and 2.
\begin{figure}[h]
\unitlength 1cm
\begin{picture}(15,5)
\epsfxsize 20cm 
\put(-2,-18.5){\epsfbox{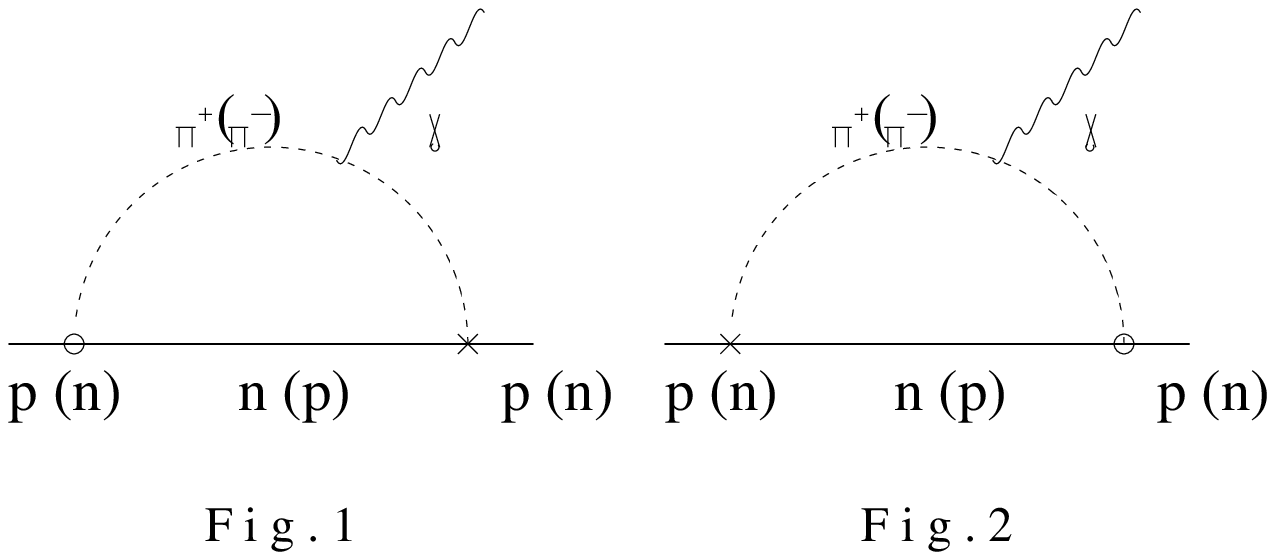}}
\end{picture}
\end{figure}
The circle on the nucleon lines refers to the usual strong interaction 
$\pi$NN vertex (coupling constant $g\sqrt{2}$), the cross describes
the P odd weak $\pi$NN interaction (coupling constant 
$\bar{g}\sqrt{2}$). The result for the nucleon AM is
\beq\label{nam}
\va_N\,=\,\va_p\,=\,\va_n\,=\,-\,{e g \bar{g} \over 12 m_p m_{\pi}}\,
\left(1-\,{6 \over \pi}\,{m_{\pi} \over m_p}\ln{m_p \over m_{\pi}}\right)
\,\mbox{\boldmath $\sigma$}.
\eeq 
The diagrams discussed lead to the same result for a proton and 
neutron since under the permutation $p \leftrightarrow n$ the strong 
coupling constant $g$ does not change, and the weak one $\bar{g}$
changes sign together with the charge $e$ of the $\pi$-meson (we
assume $e>0$, exact definitions of the strong and weak interaction 
Lagrangians and coupling constants $g$ and $\bar{g}$ are given below).
Being the only contribution to the nucleon AM, which is singular in 
$m_{\pi}$, the result (\ref{nam}) is gauge-invariant. In this respect,
it has a physical meaning.

Unfortunately, in spite of the singularity in $m_{\pi}$,
the corresponding contribution to the electron-nucleon scattering 
amplitude is small numerically as compared to other radiative corrections
to the weak scattering amplitude. Indeed, the radiative corrections to the
effective constants $C_{2p,n}$ of the proton and neutron axial 
neutral-current operators
$G / \sqrt 2\,C_{2p,n}\mbox{\boldmath $\sigma$}_{p,n}$
are~\cite{ms} 
\beq\label{cpn}
C^r_{2p}\,=\,0.032\pm 0.030\,, \quad  C^r_{2n}\,=\,-0.018\pm 0.030\,. 
\eeq
In the same units $G/\sqrt 2$, the effective axial constants induced by the 
electromagnetic interaction with
the proton and neutron anapole moments (\ref{nam}), is  
\[ C^a_{p,n}\,=\,-\,\al a_N\,(|e|G/\sqrt 2)^{-1}\,=
\,0.07\times 10^5 \bar{g}. \]
At the ``best value'' $\bar{g}= 3.3\times 10^{-7}$ (strongly
supported by the experimental result for the $^{133}$Cs anapole moment) we
obtain
\beq\label{cna}
C^a_{p,n}\,=\, 0.002.
\eeq
With this value being much less than both central points and error bars in 
(\ref{cpn}), the notion
of the nucleon AM practically has no physical meaning. This is why the 
result (\ref{nam}) was never published by the authors. It is quoted in 
book~\cite{kh} (without the logarithmic term) just as a theoretical 
curiosity. The logarithmic term in the nucleon AM is discussed in~\cite{hhm}.

However, the situation with the deuteron AM is quite different. Not only
the proton and neutron AMs add up here. The isovector part of the radiative 
corrections is much smaller than the individual contributions $C^r_{2p}$ and
$C^r_{2n}$, and is calculated with much better accuracy~\cite{ms}:
\beq\label{cdr}
C^r_{2d}\,=\,C^r_{2p}+C^r_{2n}\,=\,0.014\pm 0.003\,. 
\eeq
Moreover, there is the already mentioned, qualitatively new contribution,
due to the isovector magnetic moment of the nucleon, which dominates 
numerically the deuteron AM. Thus $a_d$ acquires a real physical meaning.

Let us go over now to the problem itself. The Lagrangians of
the strong $\pi$NN interaction and of the weak P odd one, $L_s$ and
$L_w$, respectively, are well-known:
\begin{equation}\label{s}
L_s\,=\,g\,[\,\sqrt{2}\,(\overline{p}i\gamma_{5}n \,\pi^+ 
+\overline{n}i\gamma_{5}p\, \pi^-)\,+(\,\overline{p}i\gamma_{5}p
-\overline{n}i\gamma_{5}n)\,\pi^0];
\end{equation}
\begin{equation}\label{w}
L_w\,=\,\bar{g}\,\sqrt{2}\,i\,(\,\overline{p}n \,\pi^+ 
-\overline{n}p \,\pi^-).
\end{equation}
Our convention for $\gamma_5$ is
\beq\label{g5}
\gamma_5 = \left(\begin{array}{rr}0  & -I \\
 -I &  0 \end{array}\right); 
\eeq 
the relation between our P odd $\pi$NN constant $\bar{g}$ and the 
common one $h^{(1)}_{\pi NN}$ is $\bar{g}\sqrt{2}=h^{(1)}_{\pi NN}$.

The effective nonrelativistic Hamiltonian of the P odd nucleon-nucleon 
interaction due to the pion exchange is in the momentum representation
\beq\label{q}
V(\vq)\,=\,{2 g \bar{g} \over m_p}\,{(\vI \vq) \over m_{\pi}^2 + \vq^2}
(N_1^{\dagger}\tau_{1-}N_1)\,(N_2^{\dagger}\tau_{2+}N_2).
\eeq
Here 
\[ \vI = {1 \over 2} (\mbox{\boldmath $\sigma$}_p
+\mbox{\boldmath $\sigma$}_n) \]
is the deuteron spin, 
$\vq =\p_1^{\prime}-\p_1= -(\p_2^{\prime}-\p_2)
 =\p_n^{\prime}-\p_p= -(\p_p^{\prime}-\p_n).$
Let us note that the P odd interaction (\ref{q}) which interchanges
the proton and neutron, when applied to the initial state 
$a^{\dagger}_p(\r_1)a^{\dagger}_n(\r_2)|0\rangle$ transforms it into
$a^{\dagger}_n(\r_1)a^{\dagger}_p(\r_2)|0\rangle
=-a^{\dagger}_p(\r_2)a^{\dagger}_n(\r_1)|0\rangle$. On the other hand, 
the coordinate wave function of the admixed $^3P_1$ state is proportional 
to the relative coordinate $\r$, which we define as $\r_p-\r_n$. 
Therefore, it
also changes sign under the permutation $p \leftrightarrow n$. Thus, for the 
deuteron the P odd potential can be written in the coordinate 
representation as
a simple function of $\r=\r_p-\r_n$ without any indication of the isotopic 
variables:
\beq\label{r}
V(\r)\,=\,{g \bar{g} \over 2\pi m_p}\,(-i\vI \cdot \mbox{\boldmath $\nabla$})
\,{\exp{(-m_{\pi}r)} \over r}
\eeq
The above expressions are rather standard. As standard is our sign 
convention
for the coupling constants: $g=13.45$, and $\bar{g} > 0$ for the range of
values discussed in~\cite{ddh}.

The discussed P odd interaction $V$ generates a contact current $\vj^c$. To 
obtain the explicit expression for it, we have to consider $V$ in the 
presence of
the electromagnetic field. Its including modifies the proton momentum: $\p 
\to \p - e \A$, which results in the shift $\q \to \q + e\A$ in the 
interaction (\ref{q}). Then in the momentum representation the contact 
current is
\[ \vj^c(\vq)\,=\,-\,{\partial V(\vq) \over \partial \A}\,=\,
-\,{\partial \over \partial \A}\;{2g\bar{g} \over m_p}\;
{\vI (\vq + e \A) \over m_{\pi}^2 + (\vq+ e \A)^2}\, \]
\beq\label{jc}
=\,
-\,{2 e g\bar{g} \over m_p}\;\left\{{\vI \over m_{\pi}^2 + \vq^2}\,
-\,{2 \vq (\vI \vq) \over (m_{\pi}^2 + \vq^2)^2}\right\}.
\eeq
In the last expression we have neglected the dependence of the contact 
current on $\A$. In the coordinate representation it equals
\beq\label{jcc}
\vj^c(\r)\,=\,{e g\bar{g} \over 2 \pi m_p}\,\r(\vI\mbox{\boldmath $\nabla$}) 
{e^{-m_{\pi}r} \over r}\,.
\eeq

Let us derive at first a general structure of the deuteron AM generated by 
a P odd $np$ interaction, assuming only that the deuteron is 
a pure $^3S_1$ state, bound by a spherically symmetric potential. We follow 
here essentially the line of reasoning applied in~\cite{fk}
(see also book~\cite{kh}) to the problem of a single proton in a 
spherically symmetric potential. In this case the formula for the AM
operator is~\cite{fk}
\beq\label{sao}
\va\,=\,{\pi e \over m_p}\,\{\mu_p \r \times \mbox{\boldmath $\sigma$}\,
-\,{i \over 3}\,[{\bf l}^2, \r]\}\,+\,{2\pi \over 3}\,\r \times[\r \times \vj^c], 
\eeq
with the proton magnetic moment $\mu_p=2.79$. In the case of the deuteron  
this formula generalizes to
\beq\label{dao}
\va_d\,=\,{\pi e \over 2 m_p}\,
\{\r \times (\mu_p\mbox{\boldmath $\sigma$}_p\,
-\,\mu_n\mbox{\boldmath $\sigma$}_n)\,-\,
{i \over 6}\,[{\bf l}^2, \r]\}\,+\,{\pi \over 6}\,\r \times[\r \times \vj^c],
\eeq
$\mu_n=-1.91$ is the neutron magnetic moment.
Both AM operators (\ref{sao}) and (\ref{dao}) are orthogonal to $\r$ 
(neither of them commutes with $\r$, so the orthogonality means here 
that $\va\r+\r\va=0$). 
Therefore, the contact current (\ref{jcc}) generated by the P odd pion 
exchange and directed along $\r$, does not contribute to the nuclear AM.

Let us present now the wave function of the deuteron $^3S_1$ state as
$\psi_0(r)\chi$, where $\chi$ is the spin wave function for $I=1$ (we neglect 
here and below a small $^3D_1$ admixture in the deuteron). If 
the P odd interaction conserves the total spin $\vI$ of the deuteron,
the $^3P_1$ state admixed by it can be written as $i(\vI \r/r)\psi_1(r)$ 
(both radial wave functions, $\psi_0(r)$ and
$\psi_1(r)$, are spherically symmetric). Simple calculations demonstrate that
the deuteron AM, as induced by the operator (\ref{dao}), is in the absence of
the contact contribution
\beq\label{da}
\va_d\,=\,{\pi e \over 3 m_p}\,\left(\mu_p\,-\,\mu_n \,-\,{1 \over 3}\right)\,
\int d\r r \psi_0(r) \psi_1(r).
\eeq
So, under the assumptions made, the deuteron AM should 
depend on the universal combination $\;(\mu_p-\mu_n-\,1/3)$. 

We confine mainly in our calculation to the na\"{\i}ve zero-range 
approximation (ZRA) for the deuteron wave function:
\beq\label{p00}
\psi_0^{(0)}(r)\,=\,\sqrt{{\kappa \over 2\pi}}\;{\exp{(-\kappa r)} \over r}.
\eeq
Here $\kappa=\sqrt{m_p \ep}$; $\ep = 2.23$ MeV is the deuteron binding 
energy. 

The P odd correction to the deuteron wave function due to $V(\r)$ will be 
found in the common stationary perturbation theory. In the same ZRA the admixed
$^3P_1$ states of the continuous spectrum are free. Moreover, we can choose 
plane waves as the 
intermediate states since the perturbation  $V(\r)$ selects by itself the 
$P$-state from the plane wave. Thus obtained first-order correction to the 
wave function is
\beq\label{p1}
\psi_1(\r)\,=\,\int {d \bk \over (2\pi)^3}\;
{e^{i \bk \r} \over -\ep - k^2/m_p}\;
\int d \r^{\prime}\,e^{- i \bk \r^{\prime}}
V(\r^{\prime})\, \psi_0(r^{\prime}).
\eeq
Rather lengthy calculation leads to the following expression for the matrix 
element of the radius-vector:
\beq\label{vr}
\int d \r \psi_0(r) \r \psi_1(\r)\,
=\,-\,{i \vI g \bar{g} \over 6 \pi m_{\pi}}\,{1+\xi \over (1+2\xi)^2}, 
\eeq
where $\xi = \kappa/m_{\pi}\,=\,0.32$. With this matrix element and the
operator (\ref{dao}) one obtains easily the following result for the deuteron
AM:
\beq\label{pad}
\va_d^{(0)}\,=\,-\,{e g\bar{g} \over 6 m_p m_{\pi}}
\,{1+\xi \over (1+2\xi)^2}\,\left(\mu_p-\mu_n-\,{1 \over 3}\right)\,\vI, 
\eeq  
in accordance with the general formula (\ref{da}). Our overall factor at the
structure $\;(\mu_p-\mu_n-1/3)$ is the same as that at $\;(\mu_p-\mu_n)$ in the
revised version of~\cite{ss}. However, the corresponding total result obtained 
in~\cite{ss}, even in it revised version, is not proportional to the universal 
combination $\;(\mu_p-\mu_n-\,1/3)$.  

In fact, the range $1/m_{\pi}$ of the
P odd interaction (\ref{r}) is quite comparable to the range of the usual
nuclear forces. Therefore, it is, strictly speaking, inconsistent to use the 
zero-range approximation for calculating effects induced by the perturbation 
$V(\r)$. Still, numerical estimates made with a model deuteron wave function
which has somewhat more realistic properties, indicate that the error 
introduced by
using the ZRA does not exceed 20\%. As to other sources of P violation, 
different from the pion exchange, there are no reasons to expect that in the 
case of deuteron their neglect creates a serious error if the P odd $\pi NN$
coupling constant $\bar{g}$ is at least comparable to its ``best value''.
 
It looks reasonable to combine the potential contribution (\ref{pad}) with 
the additive contribution of the nucleon anapole moments, which according 
to (\ref{nam}) is
\beq\label{nad}
\va_d^N\,=\,\va_p\,+\,\va_n\,=\,-\,{e g\bar{g} \over 6 m_p m_{\pi}}\,
\left(1-\,{6 \over \pi}\,{m_{\pi} \over m_p}\ln{m_p \over m_{\pi}}\right)
\vI.
\eeq
In this way we arrive at the final result for the deuteron AM in the 
chiral limit:
\beq\label{fff}
\va_d\,=\,-\,{e g\bar{g} \over 6 m_p m_{\pi}}\,
\left[\,0.49\,(\mu_p-\mu_n -{1 \over 3})\,+\,0.46\right]\,\vI \,
=\,-2.60\,\,{e g\bar{g} \over 6 m_p m_{\pi}}\,\vI \,.
\eeq
This result includes all contributions to the P odd amplitude of 
$ed$-scattering, which are singular in $m_{\pi}$, and thus is 
gauge-invariant, independent of the gauge choice for the Green's 
functions of heavy vector bosons. 
 
Finally, let us compare the contribution of (\ref{fff}) to the P odd $ed$ 
scattering amplitude which is due to the usual radiative corrections,  
nonsingular in $m_{\pi}$.
For the deuteron the axial operator looks as follows: 
\[ {G \over \sqrt 2}\,C_{2d}\,\vI\,. \]
The contributions to the isoscalar axial constant $C_{2d}$ originate from  
the anapole moment, from usual radiative corrections nonsingular in $m_{\pi}$, 
and from the admixture of strange quarks in nucleons~\cite{ef}. The magnitude 
of the $s$-quarks contribution is extremely interesting, but highly uncertain. 
As to the usual radiative corrections, their contribution to this constant
is found in~\cite{ms} with good accuracy (see (\ref{cdr}):
$C^r_{2d}\,=\,0.014\pm 0.003$. 
In the same units $G/\sqrt 2$ the effective axial constant induced by the 
electromagnetic interaction with the deuteron AM (\ref{fff}) is  
\beq\label{cda}
C^a_{2d}\,=\,\al a_d\,(e G/\sqrt 2)^{-1}\,=\,0.44\times 10^5 \bar{g}. 
\eeq
At the ``best value'' $\bar{g}= 3.3\times 10^{-7}$ (strongly
supported by the experimental result for the $^{133}$Cs anapole moment) we
obtain
\beq\label{ca}
C^a_{2d}\,=\,0.014 \pm 0.003.
\eeq
We use here the above estimate of 20\% for the accuracy of our 
calculation (for given $\bar{g}$). The numbers in 
(\ref{cdr}) and (\ref{ca}) are quite comparable, and taken together result 
in the following value of the total effective constant:
\beq\label{caf}
C_{2d}\,=\,C^r_{2d}+C^a_{2d}\,=\,0.028 \pm 0.005.
\eeq

We are fully aware of the extreme difficulty of the experimental
measurement of the constant $C_{2d}$. However, with such a good accuracy of 
the 
theoretical prediction (\ref{caf}), this experiment becomes a source of the
valuable information on the P odd $\pi$NN constant and on the $s$-quark
content of nucleons. As it was 15 years ago, now again ``$C_{2d}$ seems to 
be the most interesting 
parity-violating parameter accessible to atomic-physics 
experiments''~\cite{ms}, although by rather different reasons. 

Of course,
if necessary the accuracy of our prediction (\ref{ca}) can be improved by
using a more detailed and realistic description of the deuteron. On the 
other hand, the accuracy of radiative corrections (\ref{cdr}) can be also
improved, at least by using much more precise modern
experimental values of the parameters of the electroweak theory.

\bigskip

\section{The deuteron P odd, T odd moments}

The problem of the deuteron P odd, 
T odd multipoles: electric dipole, magnetic quadrupole, and the so-called
Schiff moment, was treated phenomenologically in~\cite{sfk}. Now we will
calculate the electric dipole and magnetic quadrupole moments within the 
approach applied above to the anapole. As to the Schiff moment, strong 
cancellations occur when calculating its value for the deuteron~\cite{sfk}. 
Therefore, one cannot expect reasonable accuracy for it with our ZRA 
deuteron wave function, and we will not consider here this problem.  

As distinct from the P odd, T even interaction, there are
three independent P odd, T odd effective $\pi$NN Lagrangians. They are
conveniently classified by their isotopic properties:\\
\beq\label{0}
\Delta T =0. \quad L_0\,=\,g_0\,[\,\sqrt{2}\,(\overline{p}n \,\pi^+ 
+\overline{n}p\, \pi^-)\,+(\,\overline{p}p -\overline{n}n)\,\pi^0];
\end{equation}
\begin{equation}\label{1}
|\Delta T| =1. \quad L_1\,=\,g_1\,\bar{N}N\,\pi^0\,=\,g_1\,(\,\overline{p}p \,+ 
\,\overline{n}n \,)\,\pi^0);\quad \quad \quad \quad  
\end{equation}
\[ |\Delta T| = 2. \quad L_2\,=\,g_2\,(
\,\bar{N}\mbox{\boldmath $\tau$}N\,\mbox{\boldmath $\pi$}\,-
\,3\bar{N}\tau^3 N\,\pi^0\,)
\quad \quad \quad \quad \quad \quad  \]
\beq\label{2}
\quad \quad \quad \quad
\quad \quad \quad \quad \quad \;\;\; 
=\,g_2\,[\,\sqrt{2}\,(\overline{p}n \,\pi^+\,+\,\overline{n}p\,\pi^-)\,
-\,2\,(\,\overline{p}p \,-\,\overline{n}n \,)\,\pi^0)\,].
\end{equation}
Since the possible values of the isotopic spin for two nucleons is 
$T=0,\,1$ only, the last interaction, with $|\Delta T| = 2$, is not
operative in our approach.

The effective P odd, T odd proton -- neutron interaction is derived
in the same way as in the AM problem. In the momentum representation it 
looks as follows:
\beq\label{wq}
W(\vq)\,=\,{g \over 2 m_p}\,{i\vq \over m_{\pi}^2 + \vq^2}\,
[\,(\,3g_0 - g_1\,)\,\mbox{\boldmath $\sigma$}_p\,-\,
(\,3g_0 + g_1\,)\,\mbox{\boldmath $\sigma$}_n\,].
\eeq
In the coordinate representation it is
\beq\label{ww}
W(\r)\,=\,{g \over 8\pi m_p}\,  
[\,(\,3g_0 - g_1\,)\,\mbox{\boldmath $\sigma$}_p\,-\,
(\,3g_0 + g_1\,)\,\mbox{\boldmath $\sigma$}_n\,]
\,\mbox{\boldmath $\nabla$}\,{e^{-m_{\pi}r} \over r}\,.
\eeq

The calculation of the deuteron EDM ${\bf d}_d$, 
i.e., of the $e\r_p=e\r/2$ expectation
value, goes along the same lines as that for the anapole moment and 
results in
\beq
{\bf d}\,=\,-\,{egg_1 \over 12\pi m_{\pi}}\,
{1+\xi \over (1+2\xi)^2}\,\vI.
\eeq

The magnetic quadrupole moment (MQM) operator is expressed 
through the current 
density $\vj$ as follows (see, for instance,~\cite{kh,kl}):
\beq\label{m}
M_{mn}\,=\,(r_m\ep_{nrs}+ r_n\ep_{mrs})r_r j_s.
\eeq
This expression transforms to
\beq\label{mm}
M_{mn}\,=\,{e \over 2m}\left\{\,3\mu\left[\,r_m\sigma_n + r_n\sigma_m
-\,{2 \over 3}\,(\mbox{\boldmath $\sigma$}\r)\right]\,+ 2q (r_m l_n
+ r_n l_m)\right\}.
\eeq
Here $\mu$ is the total magnetic moment of the particle, $q$ is its 
charge in the units of $e$. The magnetic quadrupole moment is the 
expectation value ${\cal M}$ of the 
operator $M_{zz}$ in the state with the maximum total angular momentum
projection $I_z=I$.

In our case, due to the spherical symmetry of
the deuteron nonperturbed wave function, the orbital contribution to $M_{mn}$
vanishes. The contact current generated by the P and T odd charged pion 
exchange, here is also directed along $\r$, and thus does not contribute to 
MQM. So, the deuteron magnetic quadrupole moment originates from the spin term 
in (\ref{mm}). It equals
\beq
{\cal M}\,=\,-\,{e g \over 12\pi m_p m_{\pi}}\,{1+\xi \over (1+2\xi)^2}\,
[\,(\,3g_0+g_1\,)\,\mu_p + (\,3g_0-g_1\,)\,\mu_n\,)\,].
\eeq

\bigskip

\begin{center}
***
\end{center}

\bigskip

We very much appreciate numerous helpful discussions with V.B. Telitsin. We
are grateful also to V.F. Dmitriev and V.V. Sokolov for the interest to the
work.
The work was supported by the Russian Foundation for Basic Research 
through Grant No. 98-02-17797, by the Ministry of Education
through Grant No. 3N-224-98, and by the Federal Program Integration-1998
through Project No. 274.

\end{document}